\documentclass[
reprint,
superscriptaddress,
amsmath,amssymb,
aps, prl
]{revtex4-2}

\usepackage{mathtools}
\usepackage{hyperref}
\usepackage[capitalize]{cleveref}
\usepackage{array}
\usepackage{enumerate}

\newcolumntype{L}{>{$}l<{$}} 
\newcolumntype{R}{>{$}r<{$}} 

\begin{document}
	\title{The asymptotically-free gauge theories}
	\author{Ben Gripaios}
	\affiliation{Cavendish Laboratory, Department of Physics, JJ Thomson Ave, Cambridge CB3 0US, UK}
	\email{gripaios@hep.phy.cam.ac.uk}
	\author{Khoi \surname{Le Nguyen Nguyen}}
	\affiliation{DAMTP, University of Cambridge, Wilberforce Road, Cambridge, CB3 0WA, United Kingdom}
	\email{kl518@cam.ac.uk}
	
	\begin{abstract}
We show how to classify the asymptotically-free gauge theories
in four spacetime dimensions, focussing here on the case of
purely fermionic matter. The classification depends on the fact (which we prove)
that both the
dimension and Dynkin index of irreducible representations of a
simple Lie algebra are strictly
increasing functions of each Dynkin label. This implies not only that the number of
asymptotically-free representations of any one semisimple Lie algebra is finite,
but also that they can be written down in a systematic fashion using
tables for the asymptotically-free irreducible representations of
simple Lie algebras, which we supply. These tables show that at most
two out of a possible ten Dynkin labels can be non-zero and that no
Dynkin label can exceed four. 
The extension to bosonic matter or
supersymmetric theories is straightforward. 
	\end{abstract}
	\maketitle
	
	Asymptotically-free (hereafter `AF') gauge theories provide
        the only known examples of interacting quantum field theories
        in four spacetime dimensions that are valid down to
        arbitrarily short scales. They play an important r\^{o}le both
        in the Standard Model and in theories that go beyond it, such
        as grand unified theories. Here we provide a first classification of
        such theories \footnote{A partial list of AF irreducible reps
          of type $A_n$ was
          given without proof in \cite{Eichten_1982}, but
         omits the 15 irreducible reps indicated with a
         $\ast$ in \cref{tab:AF_an}; a list of AF chiral irreducible reps
          of type $A_n$ and AF pseudoreal irreducible reps of $C_n$
          was given without proof in \cite{Cacciapaglia_2025}, but omits the
          irreducible rep indicated with a $\dagger$ in \cref{tab:AF_an}.}, focussing on the case
       in which any matter is fermionic; the generalization to
       theories with scalar particles or supersymmetry will become obvious.
	
Our classification relies on the following facts, of
        which only the last two are new \footnote{For the known facts,
          details can be found in
  \cite{Humphreys_1972,Freudenthal_1969}.}: 
  
	  \begin{enumerate}[(i)]
	  	\item  the gauge content of an AF gauge theory is determined by
	  	a semisimple Lie algebra; any such algebra is a sum of
	  	simple Lie algebras; the latter correspond to root systems
	  	of type $A_n$ with $n \geq 1$, $B_n$ with $n \geq 2$, $C_n$
	  	with $n \geq 3$, and $D_n$ with $n \geq 4$ (corresponding to the Lie
	  	groups $SU(n+1)$, $SO(2n+1)$, $Sp(2n)$, and $SO(2n)$,
	  	respectively), and five exceptional types $E_n$ with
                $6 \leq n \leq 8$,
	  	$F_{n}$ with $n=4$ and $G_n$ with $n=2$, where $n$ is
                the rank;	  	
	  	\item the matter content is determined by a
                  representation (hereafter `rep'), which is a sum of irreducible reps;
                  the irreducible reps of a simple Lie algebra are labelled
	  	by their highest weight
	  	$\lambda = \sum_{i=1}^n m_i \omega_i$, where the Dynkin labels $m_i$ take values
	  	in the nonnegative integers and $\omega_i$ are
	  	the fundamental weights, determined by the $n$ simple roots
	  	$\{\alpha_j\}$ via
	  	$\sloppy{2\frac{(\omega_i,\alpha_j)}{(\alpha_j,\alpha_j)} =
	  	\delta_{ij}}$, where $(,)$ is the usual real positive bilinear form
	  	induced by the Killing form, here normalized such that the
	  	highest root has length-squared two \footnote{This
                    differs from the usual normalization in the
                    physics literature, but is obviously preferable because it results
                    in an integer-valued Dynkin index.};	  	
	  	\item each simple summand of a semisimple Lie algebra has its
	  	own gauge coupling, whose one-loop beta function is proportional to 
	  	$\sloppy{-22 T(\lambda_\text{adj}) +
	  	4 T(\lambda_f) +  T(\lambda_s)}$, where $T(\lambda)$ is the
	  	nonnegative-integer-valued Dynkin index \cite{Dynkin_1957} of
	  	the rep $\lambda$ of the semisimple algebra restricted to that
	  	summand, and the terms give the
	  	contributions, respectively, of gauge bosons, Weyl
	  	fermions, and real scalar bosons;	  	
	  	\item  the Dynkin index of a rep is the sum of the Dynkin
	  	indices of its irreducible summands \footnote{For a product of
	  		reps, we instead have the formula $\sloppy{T(\lambda \otimes \lambda^\prime ) = D(\lambda)
	  			T(\lambda^\prime) + D(\lambda^\prime) T(\lambda)}$; using this, together with the fact that a
	  		product of irreducible reps contains as a summand the
	  		irreducible rep whose highest weight is the sum of the
	  		highest weights of the factors, it is possible to show that
	  		the only AF and anomaly-free chiral gauge theory that is a product of irreducible reps of a Lie algebra
	  		of type $A_n$ has $n=6$ and fermion rep $\omega_2 \otimes
	  		\omega_6$, settling a question raised in \cite{Gripaios_toappear}.} and the latter are given
	  	by Dynkin's formula $
	  	\sloppy{T(\lambda)=\frac{D(\lambda)}{D(\lambda_\text{adj})}[(\lambda+\delta,\lambda+\delta)-(\delta,\delta)]},$
	  	where $2\delta$ is the sum
	  	of the positive roots and where a rep's dimension is given by Weyl's formula $\sloppy{D(\lambda)= \frac{\prod_j
	  		(\lambda + \delta, \alpha_j)}{ \prod_{j} (\delta,\alpha_j)}},$
	  	with the products taken over the positive roots;	  	
	  	\item $T(\lambda)$ and $D(\lambda)$ are, as we will
                  show at the end, strictly-increasing functions of each
	  	Dynkin label 
                      (so $T(\lambda)$ is positive unless $\lambda$ is
                      a sum of
                      trivial reps);
\item even though the number of Dynkin labels of an AF irreducible rep of
  a simple Lie
  algebra of type $A_n, B_n, C_n,$ or $D_n$ can be arbitrarily large,
  we will show that at most two of them can be non-zero and no single
  one can exceed 4; moreover, we can choose an ordered set of fundamental weights \footnote{Our ordering convention follows that of \cite{Slansky_1981}.}
  such that, for any $n$, all but the first or last few weights
  (at most five of each)
  must vanish. 
\end{enumerate}

Crucially, fact (v) implies that for each possible gauge Lie algebra, there are
only finitely-many AF reps, once we exclude the trivial rep. Moreover, we can write them down
systematically via the following algorithm.

Firstly, we find the AF irreducible reps of each simple
summand. Because of facts (v) and (vi), these can be written down
systematically for any simple Lie algebra, by incrementing the values of the few Dynkin labels that
are allowed to be non-vanishing until the rep is no longer AF. This results in the explicit
\cref{tab:AF_an,tab:AF_bn,tab:AF_cn,tab:AF_dn,tab:AF_exceptional}. In each Table, we give the Dynkin index of the irreducible
        rep, along with the dimension (for $A_n$, we also give the
        value of the anomaly, computed using the formula in \cite{Georgi_1976}). 
        \Cref{tab:AF_an,tab:AF_bn,tab:AF_cn,tab:AF_dn} show the
        results for $A_n, B_n, C_n$, and $D_n$, respectively, while
        \cref{tab:AF_exceptional} shows the results for the
        exceptional algebras.

        	\begin{table*}
        	\caption{Non-trivial AF irreducible reps 
                  of type $A_n$, $n\geq1$, along with their dimension
                  $D$, Dynkin index $T$, anomaly $A$ (for $n \geq 2$), and
                  corresponding restrictions on the values of
                  $n$. These values of $n$ are further constrained so
                  that an irreducible rep and its dual do not both
                  appear in the table and that no irreducible rep
                  is listed twice. The adjoint rep is marked with
                  $\ddagger$. The chiral AF irreducible reps missed in
                  \cite{Eichten_1982} and \cite{Cacciapaglia_2025} are indicated with
                  $\ast$ and $\dagger$, respectively.}
        	\label{tab:AF_an}
        	\centering
        	\begin{ruledtabular}
        		\begin{tabular}{LLLLR}
        			\text{Weight} & D & T & A (n \geq 2) & n\in \\ \hline
        			
        			\rule{0pt}{4ex} \omega_1 & n+1 & 1 & 1 & \{1,...,\infty\}\\
        			
        			\rule{0pt}{4ex} \omega_2 & \frac{n(n+1)}{2} & n-1 & n-3 & \{3,...,\infty\} \\
        			
        			\rule{0pt}{4ex} \omega_3 & \frac{(n-1)n(n+1)}{6} & \frac{(n-2)(n-1)}{2} & \frac{(n-5)(n-2)}{2} & \{5,...,25\} \\
        			
        			\rule{0pt}{4ex} \omega_4 & \frac{(n-2)(n-1)n(n+1)}{24} & \frac{(n-3)(n-2)(n-1)}{6} & \frac{(n-7)(n-3)(n-2)}{6} & \{7,...,11\}\\
        			
        			\rule{0pt}{4ex} \omega_5 & \frac{(n-3)(n-2)(n-1)n(n+1)}{120} & \frac{(n-4)(n-3)(n-2)(n-1)}{24} & \frac{(n-9)(n-4)(n-3)(n-2)}{24} & \{9\} \\
        			
        			\rule{0pt}{4ex} 2\omega_1 & \frac{(n+1)(n+2)}{2} & n+3 & n+5 & \{2,...,\infty\} \\
        			
        			\rule{0pt}{4ex} 3\omega_1 & \frac{(n+1)(n+2)(n+3)}{6} & \frac{(n+3)(n+4)}{2} & \frac{(n+4)(n+7)}{2} & \{1,2^\ast...,15^\ast\} \\
        			
        			\rule{0pt}{4ex} 4\omega_1 & \frac{(n+1)(n+2)(n+3)(n+4)}{24} & \frac{(n+3)(n+4)(n+5)}{6} & \frac{(n+4)(n+5)(n+9)}{6} & \{1\} \\
        			
        			\rule{0pt}{4ex} \omega_1+\omega_2 & \frac{n(n+1)(n+2)}{3} & n^2+2n-2 & (n-2)(n+4) & \{3,...,10\} \\
        			
        			\rule{0pt}{4ex} \omega_1+\omega_3 & \frac{(n-1)n(n+1)(n+2)}{8} & \frac{(n-1)(n^2+n-4)}{2} & \frac{(n-3)(n^2+n-8)}{2} & \{4,5^{\ast,\dagger}\} \\
        			
        			\rule{0pt}{4ex} \omega_1+\omega_n\,^\ddagger & n(n+2) & 2(n+1) & 0 & \{1,... \infty\}\\
        			
        			\rule{0pt}{4ex} \omega_1+\omega_{n-1} & \frac{(n-1)(n+1)(n+2)}{2} & \frac{(n-1)(3n+4)}{2} & \frac{-n^2+5n+8}{2} & \{5,...,8\} \\
        			
        			\rule{0pt}{4ex} 2\omega_1+\omega_n & \frac{n(n+1)(n+3)}{2} & \frac{(n+3)(3n+2)}{2} & \frac{n^2+9n+6}{2} & \{2,3,4\} \\
        			
        			\rule{0pt}{4ex} 2\omega_2 & \frac{n(n+1)^2(n+2)}{12} & \frac{(n-1)(n+1)(n+3)}{3} & \frac{(n-3)(n+1)(n+5)}{3} & \{3,4,5\} \\
        			
        			\rule{0pt}{4ex} \omega_2+\omega_3 & \frac{(n-1)n(n+1)^2(n+2)}{24} & \frac{(n-1)(n+1)(5n^2+6n-24)}{24} & \frac{(n-4)(n+1)(5n^2+13n+3)}{24} & \{4\}
        		\end{tabular}
        	\end{ruledtabular}
        \end{table*}
        \begin{table}
        	\caption{Non-trivial AF irreducible reps of type $B_n$, $n\geq2$, and corresponding restrictions on the values of $n$. The adjoint rep is marked with $\ddagger$ (for $n\geq 3$) and $\ddagger\ddagger$ (for $n=2$).}
        	\label{tab:AF_bn}
        	\centering
        	\begin{ruledtabular}
        		\begin{tabular}{LLLR}
        			\text{Weight} & D & T & n\in \\ \hline
        			
        			\rule{0pt}{4ex} \omega_1 & 2n+1& 2 & \{2,\dots,\infty\} \\
        			
        			\rule{0pt}{4ex} \omega_2^\ddagger & n(2n+1)  & 2(2n-1) & \{3,\dots,\infty\} \\
        			
        			\rule{0pt}{4ex} \omega_3 & \frac{n(2n+1)(2n-1)}{3} & 2(2n-1)(n-1) & \{4,5,6\} \\
        			
        			\rule{0pt}{4ex} \omega_n & 2^n  & 2^{n-2} & \{2,\dots,9\} \\
        			
        			\rule{0pt}{4ex} 2\omega_1 & n(2n+3) & 2(2n+3) & \{2,\dots,\infty\} \\
        			
        			\rule{0pt}{4ex} \omega_1+\omega_n & 2^{n+1}n & (n+4)2^{n-1} & \{2,3,4\} \\
        			
        			\rule{0pt}{4ex} 2\omega_n^{\ddagger\ddagger} & \frac{(2n+1)!}{n!(n+1)!} &  \frac{(2n)!}{(n!)^2} & \{2,3,4\}
        		\end{tabular}
        	\end{ruledtabular}
        \end{table}
        \begin{table}
        	\caption{Non-trivial AF irreducible reps of type $C_n$, $n\geq3$, and corresponding restrictions on the values of $n$. The adjoint rep is marked with $\ddagger$.}
        	\label{tab:AF_cn}
        	\centering
        	\begin{ruledtabular}
        		\begin{tabular}{LLLR}
        			\text{Weight} & D & T & n\in \\ \hline
        			
        			\rule{0pt}{4ex} \omega_1 & 2n & 1 & \{3,...,\infty\} \\
        			
        			\rule{0pt}{4ex} \omega_2 & (n\!-\!1)(2n\!+\!1) & 2(n\!-\!1) & \{3,...,\infty\} \\
        			
        			\rule{0pt}{4ex} \omega_3 & \frac{2(n\!-\!2)n(2n\!+\!1)}{3} & (2n\!-\!1)(n\!-\!2) & \{3,\dots,8\} \\
        			
        			\rule{0pt}{4ex} \omega_4 & \frac{(n\!-\!3)n(2n\!-\!1)(2n\!+\!1)}{6} & \frac{2(n\!-\!3)(n\!-\!1)(2n\!-\!1)}{3} & \{4,5\} \\
        			
        			\rule{0pt}{4ex} \omega_5 & \frac{(n\!-\!4)(n\!-\!1)n(2n\!-\!1)(2n\!+\!1)}{15} & \frac{(n\!-\!4)(n\!-\!1)(2n\!-\!1)(2n\!-\!3)}{6} & \{5\} \\
        			
        			\rule{0pt}{4ex} 2\omega_1^\ddagger & n(2n\!+\!1) & 2(n\!+\!1) & \{3,...,\infty\} \\
        			
        			\rule{0pt}{4ex} 3\omega_1 & \frac{2n(n\!+\!1)(2n\!+\!1)}{3} & (n\!+\!1)(2n\!+\!3) & \{3\} \\
        			
        			\rule{0pt}{4ex} \omega_1+\omega_2 & \frac{8(n\!-\!1)n(n\!+\!1)}{3} & 4(n\!-\!1)(n\!+\!1) & \{3\} \\
        			
        			\rule{0pt}{4ex} \omega_1+\omega_3 & \frac{(n\!-\!2)(n\!+\!1)(2n\!-\!1)(2n\!+\!1)}{2} & \scriptstyle{2(n\!-\!2)(n\!+\!1)(2n\!-\!1)} & \{3\}
        		\end{tabular}
        	\end{ruledtabular}
        \end{table}
        \begin{table}
        	\caption{Non-trivial AF irreducible reps of type $D_n$, $n\geq4$, and corresponding restrictions on the values of $n$. The adjoint rep is marked with $\ddagger$.}
        	\label{tab:AF_dn}
        	\centering
        	\begin{ruledtabular}
        		\begin{tabular}{LLLR}
        			\text{Weight} & D & T_2 & n\in \\ \hline
        			
        			\rule{0pt}{4ex} \omega_1 & 2n & 2 & \{4,...,\infty\} \\
        			
        			\rule{0pt}{4ex} \omega_2^\ddagger & n(2n-1)  & 4(n-1) & \{4,...,\infty\} \\
        			
        			\rule{0pt}{4ex} \omega_3 & \frac{2(2n-1)(n-1)n}{3} & 2(n\!-\!1)(2n\!-\!3) & \{5,6\} \\
        			
        			\rule{0pt}{4ex} \omega_{n-1},\omega_n & 2^{n-1} & 2^{n-3} & \{4,...,10\} \\
        			
        			\rule{0pt}{4ex} 2\omega_1 & (2n-1)(n+1) & 4(n+1) & \{4,...,\infty\} \\
        			
        			\rule{0pt}{4ex} \omega_1\!+\!\omega_{n-1},\omega_1\!+\!\omega_n & (2n-1)2^{n-1} & (2n+7)2^{n-3} & \{4,5\} \\
        			
        			\rule{0pt}{4ex} \omega_{n-1}+\omega_n & \frac{n.2^n (2n-1)!!}{(n+1)!} & \frac{(n-1)2^n(2n-3)!!}{n!} & \{4\} \\
        			
        			\rule{0pt}{4ex} 2\omega_{n-1},2\omega_n & \frac{2^{n-1}(2n-1)!!}{n!} & \frac{2^{n-1}(2n-3)!!}{(n-1)!} & \{4,5\} 
        		\end{tabular}
        	\end{ruledtabular}
        \end{table}
        \begin{table}
        	\caption{Non-trivial AF irreducible reps of the exceptional Lie algebras. The adjoint rep is marked with $\ddagger$.}
        	\label{tab:AF_exceptional}
        	\centering
        	\begin{ruledtabular}
        		\begin{tabular}{LLLR}
        			\text{Type} & \text{Weight} & D & T \\ \hline
        			\rule{0pt}{4ex} G_2 &  \omega_1^\ddagger & 14  & 8 \\
        			& \omega_2 & 7 & 2 \\
        			& 2\omega_2 & 27  & 18 \\
        			
        			\rule{0pt}{4ex} F_4 & \omega_1^\ddagger & 52  & 18 \\
        			& \omega_4 & 26 & 6 \\
        			
        			\rule{0pt}{4ex} E_6 & \omega_1,\omega_5 & 27  & 6 \\
        			& \omega_6^\ddagger & 78 & 24 \\
        			
        			\rule{0pt}{4ex} E_7 & \omega_1^\ddagger & 133 & 36 \\
        			& \omega_6 & 56 & 12 \\
        			
        			\rule{0pt}{4ex} E_8 & \omega_7^\ddagger & 248 & 60
        		\end{tabular}
        	\end{ruledtabular}
        \end{table}
        Secondly, we find the AF reducible reps of each simple
        summand. This can be done using fact (iv) and the fact that,
        because $T$ is non-negative, we need only consider sums of the
        AF irreducible reps, which are given, along with their Dynkin
        indices, in the Tables. Here it is possible to be explicit in
        principle, but not in practice. Indeed, while we could
        list the allowed combinations of multiplicities of the
       AF irreducible reps in each of the tables, the list would be
       rather long. For $A_4$ for example, there are just 10 AF irreducible reps
       (up to conjugation), but 542\,103 AF reducible reps
       (including conjugates).

       Finally, we observe that a rep of a semisimple Lie algebra is AF if
        and only if it is AF when restricted to each of the simple
        summands. Then we use the result for the restriction that, given irreducible reps $\lambda$ and
        $\lambda^\prime$ of simple Lie algebras $\mathfrak{g}$ and
        $\mathfrak{g}^\prime$, the corresponding irreducible rep of
        $\mathfrak{g} \oplus \mathfrak{g}^\prime$ restricted to $\mathfrak{g}$ contributes
        $T(\lambda)D(\lambda^\prime)$ to the Dynkin index. Again, the fact that both $T(\lambda)$ and
        $D(\lambda)$ are strictly increasing means that one can do
        this systematically by incrementing the Dynkin labels. 

        (For
        scalar matter or theories with supersymmetry, one should
        carry out the procedure just described, but using the appropriate values of the
        coefficients in the one-loop beta function.) 

        Amusingly, there are exactly two theories with a simple Lie
        algebra and irreducible fermion rep whose
        1-loop beta function vanishes, namely the one with type $C_4$
        (corresponding to the Lie group $\mathrm{Sp} (8)$) with fermion
        rep $3\omega_1$ and the one with type $D_7$ (coresponding to the Lie group $\mathrm{SO} (14)$) with fermion rep $\omega_3$. Since in such cases the two-loop contribution to the beta function is always
        positive \cite{Caswell_1974}, these gauge theories are not AF.

       For $A_n$ with $n\geq 2$ we also need to ensure that anomalies
       cancel, at least if we want a gauge theory that is valid down to
       arbitrarily short distance scales. As \cref{tab:AF_an} shows, the only
       such gauge theory with an irreducible fermion rep for any value of $n$ is the
       obvious one, namely the (self-dual) adjoint
       rep \footnote{As \cite{Gripaios_2024} shows, there are infinitely many
           chiral anomaly-free irreducible reps for each $n \geq 4$, but
           none are AF.}. But there are
       plenty of anomaly- and asymptotically-free examples with
       reducible reps. To find all of these, one can solve the linear
       homogeneous equation corresponding to anomaly cancellation with
       coefficients given by the `$A$' column in \cref{tab:AF_an} in the
       integers (dual reps having negative the anomaly) together with
       the linear inhomogeneous inequality corresponding to asymptotic
       freedom with coefficients given by the `$T$' column in the
       nonnegative integers (dual reps having the same Dynkin index). For
       example, for $n=4$, we find that there are 10\,036
       AF and anomaly-free reducible reps, including conjugates. 
 
       We now outline how the Tables are obtained. For $A_n$, taking
       the dual of a rep sends $m_i \mapsto m_{n-i+1}$ and we include
       only one of these in the Table. The
       fundamental rep $\omega_i$ corresponds to an antisymmetric tensor of
       rank $i$ and has Dynkin index $\binom{n-1}{i-1}$
       \cite{Kumar_1997}, so increases strictly with $i$ up to
       halfway, while the
       adjoint is $\omega_1 + \omega_n$ 
       and has $T(\omega_1 +
       \omega_n)=2(n+1)$. By solving the equality $\binom{n-1}{i-1} = 11(n+1)$
       for $i$ in this interval, we find that only the fundamental reps with $i \leq 5$
       (or their duals with $n-i+1 \leq 5$) can be AF for some $n$. By fact (v), it
       follows that for any irreducible rep, only the first or last five
       Dynkin labels can be non-vanishing. This leaves just a few
       cases to check, by increasing these Dynkin labels until we find
       ourselves
       asymptotically enslaved.

       	For $B_n$, the first $n-1$ fundamental reps correspond to
        antisymmetric tensors of ascending rank with
        $\sloppy{T(\omega_i)=2\binom{2n-1}{i-1}}$ \cite{Kumar_1997}, which strictly increases with $i$,
        and the last fundamental rep corresponds to the spinor with
        $T(\omega_n)=2^{n-2}$, while the adjoint rep is $\omega_2$ for
        $n\geq3$ and $2\omega_2$ for $n=2$, with Dynkin index
        $T(\lambda_\text{adj})=2(2n-1)$. Now the only possibly AF fundamental
        reps for some $n$ are $\omega_{1,2,3,n}$.

	For $C_n$, the fundamental rep $\omega_i$ corresponds to a
        traceless antisymmetric tensor of rank $i$ (where traces are
        taken with the Levi-Civita tensor $\epsilon_{ij}$)
        with Dynkin index $\binom{2n-2}{i-1}-\binom{2n-2}{i-3}$ \cite{Kumar_1997}, which
        strictly increases up to $i =
        \lceil\frac{1}{2}(1+2n-\sqrt{1+2n})\rceil$ and decreases afterwards, while
        the adjoint rep is $2\omega_1$ with Dynkin index $2(n+1)$. The
        upshot is that only the fundamental reps with $i\leq 5$ can be
        AF for some $n$.

        	For $D_n$, the first $n-2$ fundamental reps correspond
                to antisymmetric tensors of ascending rank with
                $\sloppy{T(\omega_i) =\binom{2n-2}{i-1}}$ \cite{Kumar_1997}, which is strictly
                increasing, and the last 2 correspond to 2 spinor
                reps, both with Dynkin index
                $2^{n-3}$. (When $n$ is even the spinor reps
                are self-conjugate, while for $n$ odd they are
                conjugate to one another.) The adjoint rep is
                $\omega_2$ with Dynkin index $4(n-1)$. The upshot is
                that the only values of $i$ we need consider are
                the first three and the last two.

For the exceptional algebras, there are anyway only a finite number of
cases to check, so we hit them with a sledgehammer. 
	
We finish by proving that $D(\lambda)$ and $T(\lambda)$ are
        strictly increasing functions of the Dynkin labels. Starting
        with $D(\lambda)$, we need only show strict monotonicity as it
        is manifestly positive (and integral). In Weyl's formula,
        $D(\lambda)$ depends on the Dynkin labels $m_i$ via a product
        of factors $(\lambda + \delta,\alpha_j)$, with $\alpha_j$ a
        positive root. Increasing any $m_i$ by 1 shifts this factor by
        $(\omega_i, \alpha_j)$, as $(,)$ is linear. By the definition
        of $\omega_i$, each such shift is nonnegative, and at least
        one is positive. So the numerator strictly increases {\em
          ergo} the dimension is strictly monotone {\em ergo} strictly
        increasing. For $T(\lambda)$, the prefactor in Dynkin's
        formula is positive, so it suffices to show that $(\lambda,\lambda) + 2(\lambda,\delta)$ strictly increases. Increasing any $m_i$ by 1 shifts it by $(\omega_i,\omega_i)+2(\delta,\omega_i)+2(\lambda,\omega_i)$, which is positive as the first two terms are positive and the third
is nonnegative. The desired result follows. It follows too
that the quadratic Casimir is strictly increasing, but we have no
need of that here.

	{\em Acknowledgements.}
	This work was partially supported by STFC consolidated grant
        ST/X000664/1 and a Trinity-Henry Barlow Scholarship.
	
	\bibliography{asymp_references}  
\end{document}